\documentclass[conference]{IEEEtran}
\IEEEoverridecommandlockouts
\usepackage{cite}
\usepackage{hyperref}
\usepackage{amsmath,amssymb,amsfonts}
\usepackage{algorithmic}
\usepackage{graphicx}
\usepackage{textcomp}
\usepackage{xcolor}
\def\BibTeX{{\rm B\kern-.05em{\sc i\kern-.025em b}\kern-.08em
    T\kern-.1667em\lower.7ex\hbox{E}\kern-.125emX}}

\definecolor{contentclr}{rgb}{0.1,0.3125,0.54}
\definecolor{urlclr}{rgb}{0,0.254,0.426}
\definecolor{citeclr}{rgb}{0.125,0.414,0.363}
\hypersetup{
    colorlinks=true,
    linkcolor=contentclr,
    urlcolor=urlclr,
    citecolor=citeclr
}
\usepackage{array}
\usepackage{multirow}
\newcolumntype{L}[1]{>{\raggedright\let\newline\\\arraybackslash\hspace{0pt}}m{#1}}
\newcolumntype{C}[1]{>{\centering\let\newline\\\arraybackslash\hspace{0pt}}m{#1}}
\newcolumntype{R}[1]{>{\raggedleft\let\newline\\\arraybackslash\hspace{0pt}}m{#1}}
    
\begin{document}

\title{An Integrated Genomics Workflow Tool: Simulating Reads, Evaluating Read Alignments, and Optimizing Variant Calling Algorithms}

\author{
\IEEEauthorblockN{1\textsuperscript{st} Fathima Nuzla Ismail}
\IEEEauthorblockA{\textit{Information Science} \\
\textit{University of Otago}\\
Otago, New Zealand \\
fathima.nuzla.ismail@gmail.com \\
0000-0002-0716-1478}
~\\
\and
\IEEEauthorblockN{2\textsuperscript{nd} Shanika Amarasoma*}
\IEEEauthorblockA{\textit{AI and Advanced Analytics} \\
\textit{Independent Researcher}\\
Colombo, Sri Lanka \\
shanika.amarasoma@gmail.com \\
0000-0001-9509-0069}
*Corresponding author
}

\maketitle

\begin{abstract}
Next-generation sequencing (NGS) is a pivotal technique in genome sequencing due to its high throughput, rapid results, cost-effectiveness, and enhanced accuracy. Its significance extends across various domains, playing a crucial role in identifying genetic variations and exploring genomic complexity. NGS finds applications in diverse fields such as clinical genomics, comparative genomics, functional genomics, and metagenomics, contributing substantially to advancements in research, medicine, and scientific disciplines. Within the sphere of genomics data science, the execution of read simulation, mapping, and variant calling holds paramount importance for obtaining precise and dependable results. Given the plethora of tools available for these purposes, each employing distinct methodologies and options, a nuanced understanding of their intricacies becomes imperative for optimization. This research, situated at the intersection of data science and genomics, involves a meticulous assessment of various tools, elucidating their individual strengths and weaknesses through rigorous experimentation and analysis. This comprehensive evaluation has enabled the researchers to pinpoint the most accurate tools, reinforcing the alignment between the established workflow and the demonstrated efficacy of specific tools in the context of genomics data analysis. To meet these requirements, ``VarFind'', an open-source and freely accessible pipeline tool designed to automate the entire process has been introduced (VarFind GitHub repository: \href{https://github.com/shanikawm/varfinder}{VarFinder}.)
\end{abstract}


\section{Introduction}
Next-generation sequencing (NGS) is crucial in genome sequencing because it offers high throughput, rapid results, cost-effectiveness, and improved accuracy \cite{Smith2020}. It plays a vital role in identifying genetic variations and exploring genomic complexity and has widespread applications in clinical genomics, comparative genomics, functional genomics, and metagenomics \cite{Brown2017}. NGS transforms the understanding of genetics, impacting research, medicine, and various scientific fields. NGS aids in personalized medicine by identifying genetic variations unique to individuals, which are crucial for assessing disease risks and selecting appropriate treatments \cite{Vegter2018}. NGS has transformed diagnostics within clinical settings, allowing for precise molecular profiling and ultimately enhancing patient care outcomes \cite{Rehm2017}. Moreover, NGS is pivotal in advancing functional genomics research by uncovering gene expression patterns and regulatory mechanisms. Its utility extends to environmental and agricultural studies, where it facilitates microbial community analysis and improves crop breeding strategies \cite{Esposito2016}. NGS has become essential for advancing biological research, medicine, and biotechnology. 

However, Next-generation sequencing (NGS) confronts a few potential limitations. Despite their low error rates, inaccuracies during library preparation, sequencing chemistry, and base calling can compromise data accuracy, particularly in detecting rare genetic variations. Sequencing coverage bias in regions characterized by extreme GC content complicates the accurate assessment of genetic variances. Substandard data quality may arise from sample impurities or degradation, underscoring the need for rigorous sample preparation protocols and quality control measures. Moreover, the substantial volume of NGS data strains computational resources. At the same time, the expenses associated with sequencing instruments, reagents, and analysis may need to be revised to ensure the feasibility of large-scale projects. Ethical concerns regarding genetic data privacy and responsible data management are also paramount, particularly within clinical contexts \cite{Goodwin2016}.

The next step is Downstream analysis in amplicon sequencing, which involves a series of computational and analytical steps applied to the generated data. This process includes quality control, data preprocessing, sequence alignment (if applicable), identification of taxonomic information or genetic variants, diversity assessment, statistical analysis, visualization, functional annotation, sample clustering, and, most importantly, deriving meaningful biological insights from the data \cite{Johnson2021}. These steps help researchers extract valuable information and draw conclusions from amplicon sequencing experiments, making it a crucial phase in the research process. In this process, identifying the difference between the target and reference datasets is crucial for the prediction outcomes, which involves treating the errors and processing various tools for different steps. Once the required variant is determined, it is convenient to address the medical prediction. However, downstream analysis requires several tools and complex processes in using all these tools to address a specific task, as each tool has a different input and output format. In a standard process, it is essential for the subsequent analysis to be both rapid and consistent. To address all these requirements, the researchers have developed VarFind, an open-source, free pipeline tool \href{https://github.com/shanikawm/varfinder}{VarFinder} to automate the process.  

\begin{figure*}
\centering
\includegraphics[trim=0 4cm 0 0, width=\textwidth]{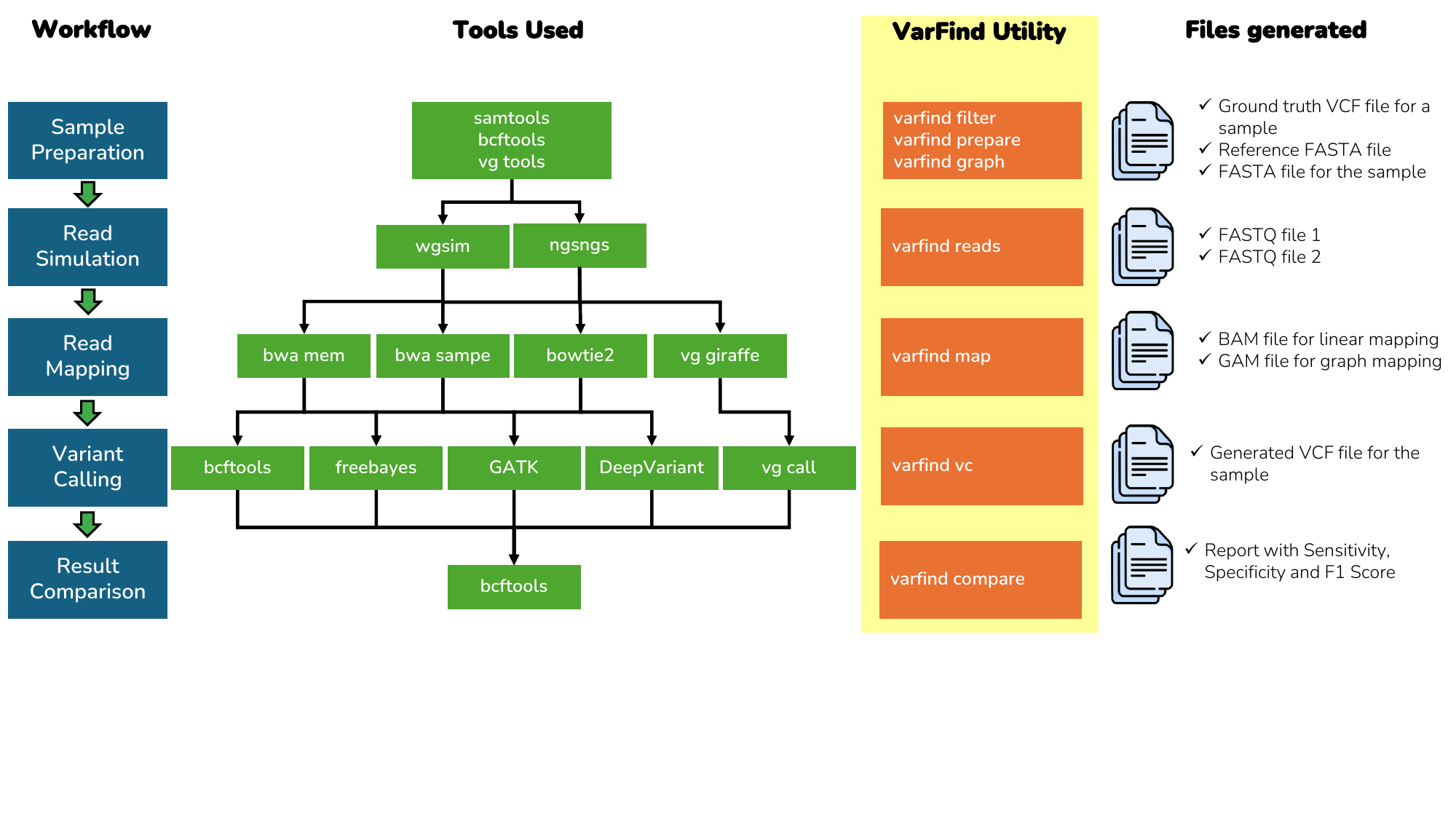}
\caption{Workflow diagram with tools used and files generated in each step}
\label{fig:workflow}
\end{figure*}

\begin{enumerate}
    \item varfind filter (Filter and index a chromosome from a fasta file)
    \item varfind prepare(creating ground truth data files), 
    \item varfind reads (simulating reads using different tools available)
    \item varfind map (align reads to a reference using different methods)
    \item varfind vc (variant calling using different methods)
    \item varfind compare (comparing the output with ground truth data)
\end{enumerate}
VarFind requires minimal computational skills, as most complex commands are automated as a script. VarFind guarantees that the processed data from sequencing undergoes a consistent and uniform treatment, promoting the ease of reproducing results and making comparisons across various samples more straightforward. The user's input requirements are significantly streamlined, encompassing just two input files fasta and reference vcf. The ``.fasta'' or ``.fa'' file is a plain text document containing one or more sequences. Each sequence typically consists of two parts: a header line, starting with ``$>$'', which provides a brief description or identifier for the sequence, and sequence data, which comprises a series of letters representing nucleotides (e.g., A, T, G, C for DNA) or amino acids (using one-letter codes for proteins). The sequence data is presented as a continuous string of characters. A Variant Call Format (VCF) file is a standard format in genomics used to store information about genetic variations in DNA or RNA sequences. It includes details about the variants' genomic positions, allele types, quality scores, and genotype data for multiple samples. VCF files are crucial for genetic variant analysis in fields like genetics, genomics, and personalized medicine. They consist of a header section with metadata and data rows representing individual variants, making them essential for variant discovery and interpretation.  

\section{Materials and Methods}

The GRCh38-related VCF files and FSTA files are used for the experiment. GRCh38, or Genome Reference Consortium human genome build 38, is the latest version of the human reference genome, released in December 2013 \cite{grch382022}. It offers significant improvements over its predecessor, GRCh37, by addressing gaps and inaccuracies and incorporating alternative loci to represent genetic diversity comprehensively. Notably, it includes patch sequences for updates without altering the primary reference. GRCh38 is the standard for genomic research, ensuring consistency and compatibility across various applications and bioinformatics tools. The Genome Reference Consortium continues to update and refine this essential genomics and genetics research resource.
The 1000 Genomes Project conducted extensive sequencing, generating over 100 trillion basepairs of short-read sequence data from more than 2600 samples across 26 populations over five years. In its final phase, the project provided more than 85 million genotyped and phased variants using the human reference genome assembly GRCh37. Although an updated reference assembly, GRCh38, was released in late 2022 \cite{grch382022}, the project needed time constraints to transition to it.

While it's possible to adapt the project's variants to GRCh38 by coordinate remapping, this process is potentially error-prone and limited to non-repetitive and unchanged regions between the two assemblies. Additionally, it would miss variants in newly added GRCh38 regions. To ensure the highest quality variants and genotypes on GRCh38, the best approach is to re-align the reads and re-call the variants based on the new alignment.

As the initial step in the variant calling for the 1000 Genomes Project data, the project has completed remapping all sequence reads to GRCh38 using scaffold-aware BWA-MEM. The resulting alignments are accessible in the CRAM format, a reference-based sequence compression format. Researchers can access this data on the project's FTP site and the European Nucleotide Archive to facilitate variant discovery on both the primary sequences and alternative contigs of GRCh38.

\subsection{varfind filter}
The selected fasta file consists of multiple chromosomes. However, for this experiment, we focus only on the chromosome 20. These varfind filter steps will filter chromosome 20 and create an index. The \texttt{samtools} was used for this task. 

\subsection{varfind prepare}
GRCh38 vcf file consists of 2000+ samples. We have selected a single sample to do this experiment. So, this step will filter a given sample from the vcf file and create the sequence for that sample using the reference sequence. Both \texttt{samtools} and \texttt{bcftools} are used for this purpose.

\subsection{varfind reads}
In this step, we simulate reads from the fasta file, which is generated in the varfind prepare step. Genomic read simulation is a computational technique used in genomics and bioinformatics to create artificial DNA or RNA sequencing data that replicates real-world sequencing experiments. This simulation generates sequences of DNA or RNA bases, mimics read length distribution, incorporates errors and variations, and allows control over coverage and depth. It helps researchers assess and validate bioinformatics algorithms, benchmark tools, train bioinformaticians, and conserve resources by reducing the need for extensive experimental sequencing. Genomic read simulation is valuable for improving genomics research and analysis workflows. In this process, the researchers have been able to specify the commands using a read simulation tool (either \texttt{wgsim} or \texttt{ngsnsgs}) with the required read length and coverage. This step will generate pared read FASTQ files at the end. A FASTQ file is a standard bioinformatics format used to store DNA or RNA sequencing data, including the nucleotide sequences and their associated quality scores. It is essential for representing raw sequencing data generated by high-throughput sequencing technologies. FASTQ files contain sequence data, quality scores, and have a specific format with four lines per read entry. These files can also handle paired-end reads, where two reads come from the same DNA fragment. FASTQ files play a vital role in genomics analyses, serving as input for tasks like read Alignment, variant calling, and gene expression quantification.

Read coverage in genomics refers to the average number of times a specific genomic position is sequenced by DNA or RNA fragments (reads) during a sequencing experiment. It is a crucial metric for assessing the depth and quality of sequencing data. Adequate coverage is essential for accurate variant calling, gene expression analysis, and genome assembly. Coverage is influenced by sequencing depth, library preparation, and target regions, and researchers often set minimum coverage thresholds based on their analysis goals. Uniform coverage across the genome is also vital for reliable genomic analysis. Visualization tools help researchers assess coverage patterns in the data.

\subsubsection{wgsim}
\texttt{wgsim} is a bioinformatics tool within the SAMtools \cite{Samtools2021} used to simulate DNA sequencing reads from a reference genome. It allows researchers to create synthetic sequencing data with specific characteristics, including read length, quality scores, and error rates. This simulated data is valuable for benchmarking bioinformatics tools, training, and generating customized datasets for various genomics research applications. \texttt{wgsim} is a command-line tool that offers flexibility in simulating both single-end and paired-end reads, making it a valuable resource in genomics and sequencing data analysis.

\subsubsection{ngsngs}
This \texttt{ngsngs} \cite{ngsngs2023} read simulator outperforms existing methods and software in terms of speed, allowing for the rapid generation of simulated reads. NGSNGS offers the capability to simulate reads that closely resemble the characteristics of specific sequencing platforms, taking into account nucleotide quality score profiles. Additionally, it incorporates a post-mortem damage model, making it particularly suitable for simulating ancient DNA samples. The simulated sequences are derived from a reference DNA genome with the option of replacement. This reference can represent various scenarios, including a haploid genome, polyploid assemblies, or population haplotypes. Notably, NGSNGS enables users to directly simulate known variable sites. The program is implemented within a multithreading framework, resulting in significantly faster execution times than existing tools. Furthermore, it expands the range of available features and output formats.

\subsection{varfind map}
The next step is to align the read to the reference. Reads Alignment is the process of mapping short DNA or RNA sequences (reads) to a reference genome or transcriptome. Its primary purpose is to determine where each read originates in the reference, which is crucial for downstream genomics analyses like variant calling and gene expression quantification. Alignment considers criteria like sequence similarity, handles sequencing errors, and accommodates paired-end reads. It faces challenges in repetitive regions and structural variations. Alignment tools, such as BWA, Bowtie, or graph-based approach, play a vital role in this process, providing foundational data for interpreting sequencing experiments and understanding genetic information. The output file of this step will be a BAM file, which is the binary format of the SAM in the case of linear mapping, and a GAM file in the case of graph-based mapping. 

A SAM file, or ``Sequence Alignment/Map'' \cite{SAMfile2009}, is a text-based format in genomics for storing information about DNA or RNA sequencing reads that have been aligned to a reference genome or transcriptome. SAM files include details such as alignment positions, sequences, and quality scores, and they begin with a metadata header section. They are human-readable, making them useful for data inspection and sharing. SAM files are employed in variant calling and genomics analysis. While versatile, they can be relatively large, leading to the use of binary BAM files for efficient storage and processing in large-scale sequencing projects. SAM files are compatible with various bioinformatics tools and are valuable for data visualization and manual inspection.

\subsubsection{BWA MEM}
``BWA-MEM'' (Burrows-Wheeler Aligner - Maximal Exact Matches) \cite{bwamem2013} is a widely used bioinformatics tool for the Alignment of DNA sequencing reads to a reference genome. It's optimized for longer reads, can handle paired-end data, and is capable of aligning reads with insertions or deletions (indels). ``bwa mem'' calculates mapping quality scores for each Alignment, ensuring high accuracy. Users provide the reference genome and sequencing data as input, and it outputs alignments in SAM or BAM format. Known for its speed and accuracy, ``bwa mem'' is a key tool in genomics for various analyses, including variant calling and gene expression quantification.

\subsubsection{BWA SAMPE}
``BWA SAMPE'' is a command-line tool included in the Burrows-Wheeler Aligner (BWA) software package \cite{bwa2009}, a widely used bioinformatics tool for aligning DNA sequencing reads to a reference genome. Specifically, ``bwa sampe'' is used to perform a paired-end alignment of sequencing reads. Paired-end sequencing generates two reads for each DNA fragment, one from each end of the fragment. ``bwa sampe'' aligns these pairs of reads together, taking into account their relative positions and orientations to provide accurate alignment results.

\subsubsection{BOWTIE2}
Bowtie2 is a versatile and widely used bioinformatics tool for aligning DNA or RNA sequencing reads to a reference genome \cite{Langmead2012-gb}. It utilizes the Burrows-Wheeler Transform (BWT) algorithm for fast and memory-efficient Alignment. Key features include support for paired-end sequencing data, gap alignment, consideration of quality scores, multithreading for speed, and output in standard formats like SAM and BAM. Bowtie 2 is used in various genomics applications and is known for its speed, accuracy, and compatibility with large-scale sequencing datasets

\subsubsection{VG GIRAFFE}
Giraffe is a mapping tool designed to align short sequencing reads to haplotypes, resulting in alignments integrated within a sequence graph \cite{Hickey2020-dx}. This approach leverages the observation that the majority of errors in Illumina sequencing data involve base substitutions. Additionally, it assumes that common insertions and deletions (indels) are already present in the haplotypes. As a result, the tool attempts to align reads without introducing gaps before considering dynamic programming-based alignment. Giraffe utilizes the GBWT index, a compressed self-indexed representation capable of handling numerous haplotypes within a graph. This index recasts the graph as an alignment of these haplotypes. The graph defines equivalent positions in the haplotypes, while the haplotypes specify which paths in the graph are relevant. Giraffe operates in graph coordinates, mapping reads to the graph, with a focus on paths consistent with the known haplotypes. This approach proves effective, particularly in complex graph regions where the number of potential paths may be extensive, but the majority of them represent rare or non-existent sequences.

\subsection{varfind call}
The most important step is finding variants in this process. Variant calling is a fundamental process in genomics that involves identifying and characterizing genetic variations within an individual's genome or transcriptome compared to a reference sequence. This process is crucial for understanding genetic diversity, conducting disease association studies, and enabling clinical genomics. It begins with sequencing data, aligns reads to a reference genome, detects variants like SNPs and indels, applies quality control, annotates variants, assigns genotypes, and represents results in standard formats. Variant calling is vital in various fields, from basic research to clinical applications.

\subsubsection{bcftool call}
\texttt{bcftools call} is a command-line tool included in the BCFtools software package \cite{Samtools2021}, which is commonly used in bioinformatics and genomics for manipulating and analyzing variant call format (VCF) files. Specifically, \texttt{bcftools call} is used for variant calling, which identifies genetic variants, such as single nucleotide polymorphisms (SNPs) and insertions/deletions (indels), in sequencing data. It can produce variant calls and genotypes in various output formats, including VCF, BCF (Binary Call Format), and other tab-delimited formats. 

\subsubsection{FreeBayes}
\texttt{freebayes} is an open-source bioinformatics tool used to discover genetic variants in DNA or RNA sequencing data, such as SNPs, indels, and structural variations \cite{FreeBayes2012}. It employs a Bayesian statistical framework to estimate variant probabilities based on sequencing quality and read depth. FreeBayes supports multi-sample analysis, excels in indel calling, assesses allele frequencies, and provides options for quality filtering. The tool outputs results in standard VCF format and is widely used in genomics research for variant discovery, population genetics, and disease variant identification.

\subsubsection{GATK HaplotypeCaller}
\texttt{GATK HaplotypeCaller} is a bioinformatics tool used in DNA sequencing data for variant calling \cite{gatk2018}. It stands out for its local de novo assembly approach, active region determination, and haplotype-based variant calling, which enhance the accuracy of variant detection, particularly for indels. The tool assigns genotypes, supports joint variant calling for multiple samples, offers quality filters, and produces output in variant call format (VCF). It is widely used in genomics research for accurate variant identification in applications such as disease variant discovery and population genetics studies.

\subsubsection{DeepVariant}
\texttt{DeepVariant} is a bioinformatics tool that utilizes deep learning, specifically Convolutional Neural Networks (CNNs), to accurately identify genetic variants in DNA sequencing data \cite{DeepV2018}. It can detect single nucleotide polymorphisms (SNPs) and insertions/deletions (indels) and perform joint calling for multiple samples. \texttt{DeepVariant} produces variant calls in standard VCF format and is known for its high accuracy, making it valuable for genomics research in disease variant discovery and population genetics. It is an open-source tool with active community support.

\subsubsection{vg call}
\texttt{vg call} is typically used for variant calling, a process in genomics that involves identifying genetic variants, such as single nucleotide polymorphisms (SNPs) and insertions/deletions (indels), in sequencing data when a variation graph is used as the reference instead of a traditional linear genome \cite{Hickey2020-dx}. Instead of aligning sequencing reads to a linear reference genome, \texttt{vg call} operates on variation graphs. These graphs represent the reference genome and genetic variants, allowing for a more flexible and accurate representation of gene diversity.

\subsection{varfind compare}
In this step, the ground truth VCF file and generated VCF file with \texttt{varfind call} are compared using \texttt{bcftools isec}. This command will produce a report with 22 parameters. (The parameters are Run Time, Ground Truth SNPs, Ground Truth INDELs, Identified SNPs, Identified INDELs, Private SNPs, Private INDELs, Matched SNPs,	Matched INDELs,	 TP,	 FP,	 TN,	 FN,	 SNP Sensitivity, SNP Specificity,	 SNP F1 Score, INDEL Sensitivity, INDEL Specificity, INDEL F1 Score,	Overall Sensitivity, Overall Specificity, and Overall F1 Score).

The performance was compared based on workflow runtime, the number of ground truth SNPs detected, and the number of ground truth INDELs caught. To determine sensitivity and specificity, we initiated the process by identifying true positive (TP), true negative (TN), false positive (FP), and false negative (FN) variants.

\begin{table*}
\centering
\caption{Average statistics of 10 samples (with average 1,880 SNPs and 300 INDELs) for the 26 workflows \label{tab:stats}}

\renewcommand{\arraystretch}{1.75}
\begin{tabular}{|l|l|l|r|r|r|r|}
\hline
\textbf{Read Simulator} & \textbf{Mapper} & \textbf{Caller} & \textbf{Run Time (Sec.)} & \textbf{SNP F1 Score} & \textbf{INDEL F1 Score} & \textbf{Overall F1 Score} \\ \hline
wgsim & bwa mem & bcftools & 73 & 97.32 & 14.33 & 86.11 \\ 
wgsim & bwa mem & freebayes & 91 & 93.70 & 0.00 & 81.10 \\ 
wgsim & bwa mem & gatk HaplotypeCaller & 131 & 99.07 & 88.97 & 97.65 \\  
wgsim & bwa mem & DeepVariant & 113 & 98.24 & 88.24 & 96.85 \\ 
ngsngs & bwa mem & bcftools & 84 & 93.25 & 0.86 & 86.39 \\   
ngsngs & bwa mem & freebayes & 107 & 92.46 & 0.00 & 80.27 \\  
ngsngs & bwa mem & gatk HaplotypeCaller & 139 & 97.34 & 85.33 & 95.70 \\  
ngsngs & bwa mem & DeepVariant & 130 & 96.75 & 87.55 & 95.48 \\ \hline
wgsim & bwa sampe & bcftools & 212 & 95.63 & 14.10 & 84.87 \\  
wgsim & bwa sampe & freebayes & 227 & 91.28 & 0.00 & 79.39 \\   
wgsim & bwa sampe & gatk HaplotypeCaller & 263 & 98.52 & 88.31 & 97.10 \\  
wgsim & bwa sampe & DeepVariant & 252 & 96.92 & 87.88 & 95.68 \\ 
ngsngs & bwa sampe & bcftools & 210 & 0.81 & 5.84 & 1.63 \\  
ngsngs & bwa sampe & freebayes & 245 & 87.40 & 0.00 & 77.24 \\  
ngsngs & bwa sampe & gatk HaplotypeCaller & 279 & 95.33 & 80.71 & 93.47 \\  
ngsngs & bwa sampe & DeepVariant & 265 & 91.20 & 80.45 & 89.87 \\ \hline
wgsim & bowtie2 & bcftools & 167 & 53.53 & 11.95 & 49.83 \\   
wgsim & bowtie2 & freebayes & 201 & 20.78 & 0.00 & 19.36 \\  
wgsim & bowtie2 & gatk HaplotypeCaller & 249 & 62.77 & 67.06 & 63.28 \\  
wgsim & bowtie2 & DeepVariant & 223 & 35.96 & 45.07 & 36.88 \\ 
ngsngs & bowtie2 & bcftools & 181 & 97.12 & 12.93 & 86.66 \\  
ngsngs & bowtie2 & freebayes & 201 & 84.32 & 0.00 & 74.15 \\ 
ngsngs & bowtie2 & gatk HaplotypeCaller & 226 & 93.41 & 73.80 & 90.90 \\   
ngsngs & bowtie2 & DeepVariant & 223 & 94.39 & 83.73 & 93.00 \\  \hline
wgsim & vg giraffe & vg call & 257 & 90.03 & 90.56 & 90.09 \\ 
ngsngs & vg giraffe & vg call & 276 & 90.43 & 90.52 & 90.41 \\  \hline

\end{tabular}

\end{table*}

This was achieved by running 10 samples for 26 different workflow scenarios. Refer to Table \ref{tab:stats}, which comprises the descriptions of each workflow scenario.
The criteria for categorizing variants were as follows:

\begin{enumerate}
    \item True Positives (TP): Variants that are common and exactly matched in both VCF files. 
    \item True Negatives (TN): All non-variant nucleotides are unavailable in both VCF files.
    \item False Positives (FP): Variants identified by workflow but not available in ground truth VCF file.
    \item False Negatives (FN): Variants available in the ground truth VCF file but not detected in the workflow.
\end{enumerate}

Sensitivity and specificity were then calculated using the following formulas:
\begin{equation}
    Sensitivity = \dfrac{TP}{(TP+FN)}
\end{equation}

\begin{equation}
    Specificity  = \dfrac{TN}{(TN+FP)}
\end{equation}

The F1 score is calculated using the following formula:

\begin{equation}
    F1  = 2 \times \dfrac{Precision + Recall}{Precision \times Recall} 
\end{equation}

Where: Precision measures the proportion of true positive predictions (correctly detected SNPs) among all positive predictions made by the model. Recall, also known as sensitivity, measures the proportion of true positive predictions (correctly detected SNPs) among all actual positive instances (all true SNPs).

The methodology mentioned above was employed to assess the performance of each workflow and pipeline in variant detection. In evaluating variant detection algorithms, sensitivity, specificity, and the F1 score are fundamental metrics. Sensitivity reflects the algorithm's capacity to accurately identify true positive variants, which is particularly crucial for detecting rare genetic variations against the ground truth data. Specificity assesses the algorithm's ability to correctly identify non-variants, essential for maintaining overall accuracy, especially in datasets where common variants dominate. The F1 score balances precision (the proportion of correctly detected SNPs among all predicted positives) and recall (the ratio of correctly detected SNPs among all true SNPs). It is calculated as the harmonic mean of these two metrics, offering a single value accounting for false positives and negatives in SNP detection. To compute the F1 score, the number of true positives, false positives, and false negatives are needed. This metric is crucial for evaluating the accuracy of SNP detection methods and is commonly used in genomics research. 

\subsection{varfind pipe}
A data pipeline plays a pivotal role in bioinformatics, offering a well-organized framework for data processing and examination. It guarantees the efficient and consistent preparation of data, enabling the extraction of valuable insights or the facilitation of modelling, thereby establishing its indispensable role in contemporary data-centric research and practical applications. Here, we discuss the Varfind pipeline built using the following workflow functions. 

\begin{enumerate}
    \item Varfind reads - this workflow functionality enables different tools to simulate reads. We have used \texttt{wgsim} and \texttt{ngsngs} for this purpose. 
    \item Varfind map - this assists in aligning reads to a reference genome using various methods available. We have tested \texttt{bwa mem}, \texttt{bwa sampe}, \texttt{bowtie2} and \texttt{vg giraffe} as the methods to align reads.
    \item Varfind vc - this workflow function helps to read variant calling using various methods. Here, we used \texttt{bcftools}, \texttt{freebyes}, \texttt{GATK}, \texttt{DeepVariant} and \texttt{vg call} for variant calling.
    \item Varfind compare - One can compare the outcome with the ground truth data in this functionality. \texttt{bcftools isec} was used for this purpose. 
\end{enumerate}

Twenty-six possible workflows use different tools and methods, as shown in Figure \ref{fig:workflow}. The following section discusses the performance of each different workflow and reveals the best workflows out of the given combinations.

\subsection{Method}
When examining genetic variants, it is imperative to account for numerous potential challenges. Among these, the comprehensive assessment of the exome sequence space and its potential impact on the analysis outcomes is paramount. In the context of our study, we used the Genome assembly GRCh38 file and filtered chromosome 20 using a \texttt{varfind-filter.sh} script. This script produces the files NC\_000020.11.fa and NC\_000020.11.fa.fai.  As the next step, we prepare the sequence file and a ground truth VCF file for a random sample (HG00096) of the GRCh38 VCF file. For the experiment purpose here, we use a selected region of chromosome 30000000-32000000 for computation convenience and to reduce bias over different sequence regions. The \texttt{varfind-prepare.sh} script was then executed to prepare the sequence file HG00096.fa, index HG00096.fa.fai, and the ground truth VCF file HG00096.vcf.gz. In addition, a graph-based method is also included using the script \texttt{varfind-graphs.sh} script to prepare a reference graph using the script with five or more random samples other than the selected analysis samples. 

\begin{table*}[ht!]
\centering
\caption{10 random samples (with average 1,880 SNPs and 300 INDELs) selected for the experiment \label{tab:samples}}
\renewcommand{\arraystretch}{1.75}
\begin{tabular}{|l|l|l|l|}
\hline
\textbf{Sample Name} & \textbf{Gender} & \textbf{Nationality} & \textbf{Ancentry} \\ \hline
HG04153	&   Female		& Bengali		& South Asian Ancestry \\
HG03755	& 	Male		& Tamil		& South Asian Ancestry \\
HG02401	& 	Male		& Dai Chinese		& East Asian Ancestry \\
HG03366	& 	Female		& Esan		& African Ancestry \\
HG00150	& 	Female		& British		& European Ancestry \\
HG01961	& 	Male		& Peruvian		& American Ancestry \\
HG00590	& 	Female		& Southern Han Chinese		& East Asian Ancestry \\
HG01097	& 	Male		& Puerto Rican		& American Ancestry \\
HG00276	& 	Female		& Finnish		& European Ancestry \\
HG03259	& 	Female		& Gambian Mandinka 	& African Ancestry \\
\hline
\end{tabular}

\end{table*}

\section{Results and Discussion}

The researchers selected the region 30,000,000 to 32,000,000 of chromosome 20 of ten samples for this experiment (Table \ref{tab:samples}). They had an average total of 1,880 SNVs and 300 INDELs. Using different workflows, we estimated average SNP, INDELs and calculated the F1 score values. Table \ref{tab:stats} shows the average counts for the unique 26 workflows for all ten samples. In this study, a standardized approach was employed, wherein a uniform read length of 100 base pairs and a consistent read depth of 60 were applied across all 26 workflows. This meticulous choice aimed to eliminate potential biases associated with variations in read length and coverage depth, thus ensuring the integrity of our results.

Examining the F1 score in relation to different workflows is a critical aspect of our analysis. F1 score stands as a paramount performance metric for many tools, and it is noteworthy that the majority of these tools encounter challenges in achieving a sensitivity rate exceeding 50\%. Therefore, there is a compelling motivation to delve deeper into the impact of sequencing depth on the variant calling sensitivity of the F1 score. To address this, the respective study systematically explored many tool combinations, seeking to identify optimal pipelines, variant callers, mappers, and simulators for enhancing sensitivity. 

\subsection{Results Based on Caller}
When the results are evaluated based on the variant caller, it is evident that \texttt{GATK HaplotypeCaller} consistently outperforms the other callers in most cases, regardless of the mapper and simulator. The overall F1 Scores for \texttt{GATK} are notably high, with values recording \texttt{GATK} with \texttt{bwa mem}: 97.65382\%, \texttt{GATK} with \texttt{bwa sampe}: 97.09887\% and \texttt{GATK} with \texttt{bowtie2}: 63.27606\%. Moreover, \texttt{DeepVariant} also exhibits competitive performance, though slightly below \texttt{GATK}, with F1 Scores ranging from the high 80s to the low 90s. However, \texttt{BCFTools} and \texttt{FreeBayes} show relatively lower performance in terms of overall F1 Scores. According to the experimental outcomes, GATK HaplotypeCaller exhibits a longer runtime than alternative methods. Its execution demands significant computational resources, encompassing both memory and processing capabilities. Analyzing extensive datasets or intricate genomes may necessitate access to high-performance computing infrastructure.

\subsection{Results Based on Mapper}
Analyzing the results based on the choice of the mapper, it becomes evident that \texttt{bwa mem} consistently delivers the best results across all callers and simulators. \texttt{bwa sampe} follows as the second-best option, and \texttt{bowtie2} generally lags behind, failing to yield superior results, regardless of the caller. 

For instance, with \texttt{bwa mem}:
\begin{itemize}
    \item \texttt{bwa mem} with \texttt{GATK}: 97.65382\%
    \item \texttt{bwa mem} with DeepVariant: 96.85403\%
\end{itemize}

With \texttt{bwa sampe}:
\begin{itemize}
    \item \texttt{bwa sampe} with \texttt{GATK}: 97.09887\%
    \item \texttt{bwa sampe} with \texttt{DeepVariant}: 95.67547\%
\end{itemize}

In contrast, \texttt{bowtie2} has lower overall F1 Scores, as seen in the earlier examples. These results emphasize the importance of the choice of the mapper for achieving better variant calling results.

\subsection{Results Based on Simulator}
The results obtained with \texttt{wgsim} and \texttt{ngsngs} vary based on the caller and mapper used. This suggests that the choice of simulator may influence the variant calling performance, and further evaluation is needed to determine its effectiveness in combination with specific callers and mappers.

\subsection{Graph-Related Performance}
Notably, the overall F1 Scores across all callers and mappers consistently hover around 90\%, demonstrating the effectiveness of the variant calling tools in this context. These results reflect the high-quality output of these tools in accurately identifying and classifying variations in genomic data.

In conclusion, \texttt{GATK} stands out as the top caller performer, and \texttt{bwa mem} is the preferred mapper. When choosing project tools, these findings provide valuable insights for researchers and practitioners in the genomics field.

\section{Conclusions}

Within the nuanced landscape of genomics data science, the execution of read simulation, mapping, and variant calling emerges as an imperative undertaking for the attainment of precise and reliable results. The multitude of available tools, each with distinct methodologies and options, necessitates a meticulous understanding of their intricacies for optimal utilization. This research, positioned at the confluence of data science and genomics, is characterized by a rigorous assessment of these tools, unveiling their strengths and weaknesses through comprehensive experimentation and analysis. This discerning evaluation has empowered us to identify the most accurate tools, reaffirming the congruence between the established workflow and the demonstrated efficacy of specific tools in genomics data analysis. As a practical manifestation of the researchers' commitment to advancing this field, we introduce VarFind, an open-source and freely accessible pipeline tool designed to automate the entire process. We opted for the genomic region from 30,000,000 to 32,000,000 for ten samples on chromosome 20. The selected samples exhibited an average total of 1,880 SNVs and 300 INDELs, employing diverse workflows; we computed the mean values for SNPs and INDELs and determined the F1 score. The experiment uses the average counts of 26 distinct workflows across all 10 samples. This study adhered to a standardized methodology, ensuring consistency by utilizing a uniform read length of 100 base pairs and maintaining a constant read depth of 60 for all 26 workflows. This deliberate standardization aimed at mitigating potential biases associated with variations in read length and coverage depth, thereby upholding the reliability and robustness of our obtained results. In summary, GATK is the leading performer among variant callers, while ``BWA-MEM'' is the preferred mapper. When selecting tools for their projects, these insights hold significance for researchers and practitioners in the genomics field. Our future endeavors involve expanding this workflow framework's tools and command options. Additionally, we are actively exploring the feasibility of incorporating support for Large Language Models (LLMs), thereby replacing conventional command-line options with the capability to interpret natural language instructions and translate them into comprehensive workflow commands.

\bibliographystyle{IEEEtran}
\bibliography{main}

\end{document}